\documentclass[prx,twocolumn,superscriptaddress,showpacs,floatfix,longbibliography]{revtex4-1}%
\usepackage{amsmath,amssymb}
\usepackage{soul}

\usepackage{lmodern,adjustbox,booktabs,multirow}
\usepackage[para,online,flushleft]{threeparttable}
\usepackage[colorlinks=true,citecolor=blue]{hyperref}

\usepackage{siunitx}
\usepackage{graphicx}% Include figure files
\usepackage{dcolumn}% Align table columns on decimal point
\usepackage{bm}% bold math
\usepackage{gensymb}
\usepackage{xcolor}
\usepackage{url}
\usepackage{notes2bib}
\DeclareUnicodeCharacter{2212}{-}

\usepackage[colorlinks=true,citecolor=blue]{hyperref}

\DeclareUnicodeCharacter{2212}{-}

\begin{document}

%\preprint{APS/123-QED}

\title{Multicomponent magneto-orbital order and magneto-orbitons in monolayer VCl$_3$}

\author{Luigi Camerano}
\affiliation{Department of Physical and Chemical Sciences, University of L'Aquila, Via Vetoio, 67100 L'Aquila, Italy}

\author{Adolfo O. Fumega}
\affiliation{Department of Applied Physics, Aalto University, 02150 Espoo, Finland}

\author{Gianni Profeta}
\affiliation{Department of Physical and Chemical Sciences, University of L'Aquila, Via Vetoio, 67100 L'Aquila, Italy}
\affiliation{CNR-SPIN L'Aquila, Via Vetoio, 67100 L'Aquila, Italy}

\author{Jose L. Lado}
\affiliation{Department of Applied Physics, Aalto University, 02150 Espoo, Finland}

\begin{abstract}
Van der Waals monolayers featuring magnetic states provide a fundamental building block
for artificial quantum matter.
Here, we establish the emergence of a multicomponent ground state
featuring magneto-orbital excitations of the 3$d^2$-transition metal trihalide VCl$_3$ monolayer. 
We show that 
monolayer VCl$_3$ realizes a ground state with simultaneous magnetic and orbital ordering using density functional theory. Using first-principles methods we derive an
effective Hamiltonian with intertwined spin and orbital degree of freedom, which we demonstrate can be tuned by strain. We show that magneto-orbitons appear as the
collective modes of this complex order, and arise from coupled orbiton
magnon excitations due to the magneto-orbital coupling in the system. Our results establish VCl$_3$ as a promising 2D material to observe emergent magneto-orbital excitations and provide a platform for multicomponent symmetry breaking. 
\end{abstract}

\maketitle

Correlations in van der Waals materials represent the driving force behind a variety of 
unconventional electronic states, including unconventional
superconductivity \cite{Park2021,Cao2021,Kim2022}, 
exotic magnetism \cite{Serlin2020}, and fractional
topological states \cite{Cai2023,PhysRevX.13.031037}.
Van der Waals magnetic materials \cite{Gibertini2019,Blei2021} provide
a highly tunable platform
to engineer a variety of quantum states,
including ferromagnetism in CrBr$_3$ and CrI$_3$ \cite{Huang2017,Lado2017}, 
antiferromagnetism in FePS$_3$ \cite{Lee2016},
heavy-fermion Kondo states in CeSiI \cite{Posey2024,Fumega2024}
and in dichalcogenide bilayers \cite{Vao2021,Wan2023,Zhao2023},
quantum spin liquid candidates candidates in 
RuCl$_3$\cite{Banerjee2017,PhysRevLett.120.077203}, 
1T-TaSe$_2$ \cite{Ruan2021}
and NbSe$_2$\cite{Zhang2024},
orbital magnets
in twisted graphene bilayers \cite{Serlin2020,PhysRevB.103.035427}
and dichalcogenide bilayers \cite{PhysRevX.13.031037,PhysRevX.14.011004},
and multiferroic order in NiI$_2$ \cite{Song2022,Amini2024}
and twisted CrBr$_3$ bilayers \cite{Song2021,Xie2023,Fumega2023,PhysRevB.107.195128}.
Among them, multiferroic materials feature, besides a magnetic order in
the spin degree of freedom, an additional electronic ordering
in a spatial degree of freedom.
While multiferroic monolayers as
NiI$_2$\cite{Song2022,Fumega2022,PhysRevLett.131.036701,PhysRevB.106.035156,Amini2024,2024arXiv240816600A} feature simultaneous magnetism and
ferroelectricity, a variety of other multicomponent orders are potentially
possible, in particular associated with an
electronic reorganization
in an internal orbitals degree of freedom \cite{Tokura2000}. 

Among van der Waals magnets, 
VCl$_3$ 
provides an ideal playground for complex
electronic ordering due to the existence
of orbital degeneracy
leading to different potentially competing ground states \cite{camerano2024symmetry}. 
VCl$_3$ belongs to family of vanadium trihalide materials, magnetic insulators due to its partially filled d-shell and strong correlations \cite{mastrippolito2023polaronic,Kong:2019,Kong:2019a,Hovancik:2022,Yang:2020,DeVita:2022,Bergner:2022,mastrippolito2023polaronic,DeVita:2022}.
Orbital ordering in monolayers
provides a unique platform to stabilize exotic
multicomponent orders tunable
via substrate \cite{deng2024ferroelectricity}
electric gate \cite{Huang2018} or twist engineering,
in contrast with the more challenging control of orbital order
in bulk compounds \cite{Tokura2000,ABO3_1,ABO3_2,ABO3_3,ABO3_4}.
VCl$_3$ is an ideal
candidate for orbital order due to its weaker spin-orbit coupling, 
which often quenches orbital order in heavier
halides.

Here, we establish emergence of a multicomponent ordering in VCl$_3$, featuring
antiferro-orbital ordering coupled to a coexisting magnetic phase.
Using first principles methods, we show that the orbital degeneracy in VCl$_3$ gives
rise to different magnetic and orbital orderings, with the lowest energy
configuration tunable by an external strain.
We further show that these two orders are not independent, but strongly coupled, leading to strong magneto-orbital effects.
This coupling gives rise to the appearance of new hybrid quasiparticles
emerging from magnons and orbital excitations. Our results establish 
VCl$_3$ as a paradigmatic material to realize multiferroic orbital ordering,
providing a van der Waals monolayer enabling the observation magneto-orbital excitations. 

\begin{figure*}[ht]
        \centering
	
        \includegraphics[width=1.9\columnwidth]{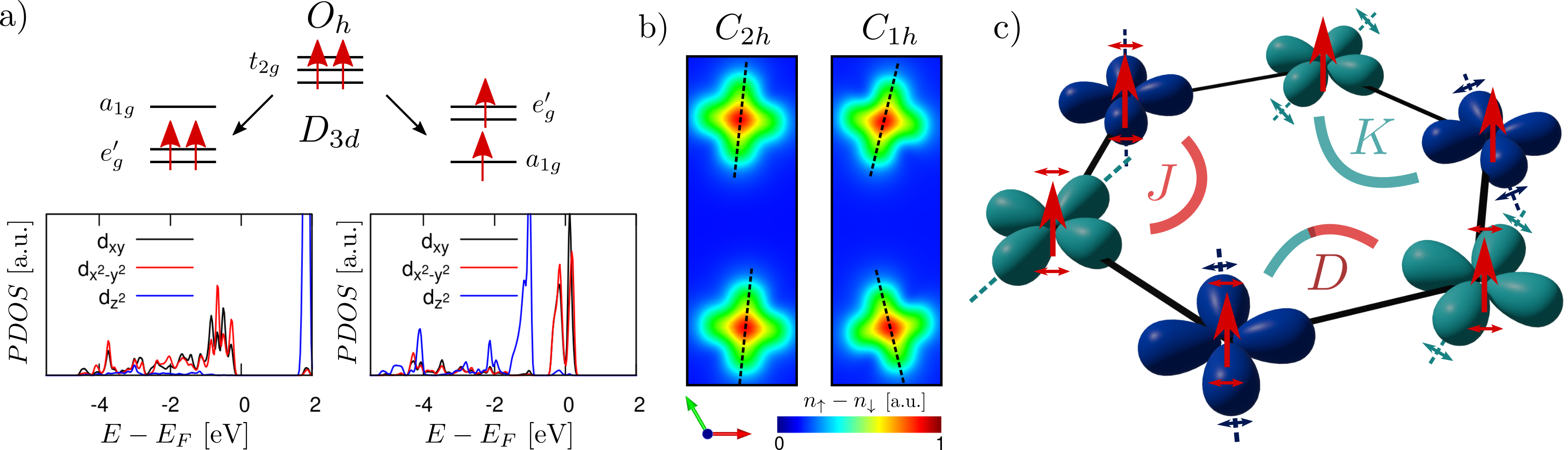}
	\caption{a) Schematic representation of the V-$d$ orbitals in a octahedral environment (O$_h$) and two possible electronic configuration after trigonal distortion (D$_{3d}$). The corresponding DFT$+U$ density of state projected on V-$d$ is reported in the lower panel showing orbital order states. b) First principle calculated magnetizatization density ($n_\uparrow-n_\downarrow$) in the case of ferromagnetic ferro-orbital (C$_{2h}$ symmetry) and ferromagnetic antiferro-orbital (C$_{1h}$ symmetry). c) Sketch of the model presented in this study, the different colors for the orbitals highlight different  V-sites in the antiferro-orbital phase. The coupling constant $J$, $K$, $D$, are referred to the model in Eq. \eqref{H}.}
	\label{Fig_1}
\end{figure*} 

We first address the electronic structure of VCl$_3$ in the monolayer limit. VCl$_3$ hosts 2 electrons in the three-fold $t_{2g}$ manifold in an octahedral environment $O_h$ (see Fig. \ref{Fig_1}a) similar to those observed in ABO$_3$ perovskites \cite{ABO3_1,ABO3_2,ABO3_3,ABO3_4}. Due to partial occupation of $d$ orbitals in $t_{2g}$ shell, a Jahn-Teller distortion \cite{jahn1937stability} of the octahedra lowers the symmetry from $O_h$ to trigonal point group D$_{3d}$ (see Fig. \ref{Fig_1}a), splitting the $t_{2g}$ manifold in a singlet $a_{1g}$ and a doublet $e'_g$. 
Spontaneous symmetry breaking leads to a further splitting of the $e'_g$ manifold lifting
the orbital degeneracy. The occupation of these two nearly-degenerate states for each V atoms gives rise to orbital ordered phases, which can have a ferro-orbital (FO) (all the V atoms are in the same orbital configuration) or anti-ferro orbital (AFO) (nearest-neighbours V atoms are in different orbital configuration) phases.
First-principles DFT$+U$ method ($U=3.2$ eV ) yield a
symmetry broken orbitally ordered phases, always favored in energy
with respect to both the $a_{1g}^1$$e'^1_g$ metallic phase and the $e'^2_g$ insulating phase. 
These orbitally ordered phases are stabilized by strong correlation, as clear from its evolution as a function of the U parameter.
As DFT calculations feature a competition between electronic and magnetic degree of freedom \cite{camerano2024symmetry},  the convergence of the orbital ordered phases in a DFT$+U$ framework is particularly challenging and requires a symmetry unconstrained unit cell and the use of the $d$-density matrix occupation control. By its suitable  initial guesses 
we compute both FO and AFO ordered phases whose magnetization densities are reported in Fig. \ref{Fig_1}b. In the case of orbital order phases the electronic instability is followed by a different lowering of the crystal symmetry (from $D_{3d}$ to $C_{2h}$ for the FO and to $C_{1h}$ for the AFO). It is worth noting how in the case of AFO the inversion symmetry of the honeycomb lattice is broken, possibly inducing a non-zero charge polarization. Comparing the energy of the FO and AFO we find that the AFO phase is favored, meaning that the ground state of the system is the AFO ferromagnetic phase. 
In order to study the stability of this solution for a wide range of structural parameters, we applied in-plane strain to the lattice, finding that the AFO phase 
always remains stable. 
The orbital ordering mechanism can be directly visualized from the density of states on vanadium $d_{z^2}$, $d_{xy}$ and $d_{x^2-y^2}$ \cite{note} showing the different occupation of the in-plane $d_{xy}$ and $d_{x^2-y^2}$ orbitals in the AFO phase but the same occupation of the $d_{z^2}$ orbital. 
The orbitally ordered phases in a magnetic van der Waals material give rise to a multicomponent ordering, which, if coupled can further stabilizes composite excitations. 
We address this possibility by deriving a model Hamiltonian from the converged first principle phases, using a spin $\vec{S}$ and pseudo-spin $\tau^z$ (describing the orbital configuration) operators to uncover the possible interaction between spin and orbital degree of freedom.

\begin{figure}
	\centering
	\includegraphics[width=0.85\columnwidth]{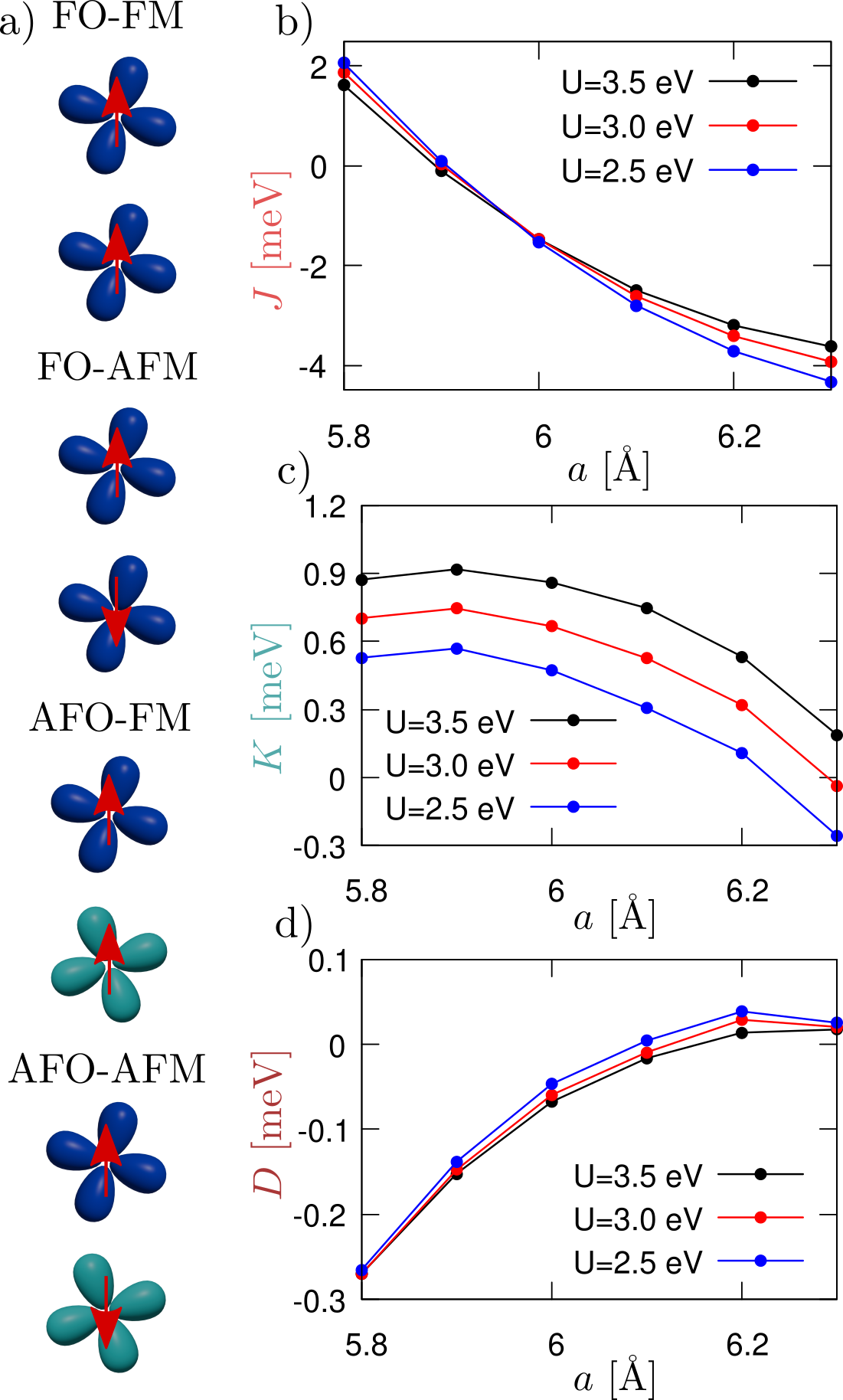}
	\caption{a) Schematic representation of the 4 states (see the main manuscript) that are needed to calculate the exchange coupling of the model. b), c) and d) Isotropic Heisenberg coupling ($J$), anisotropic orbital exchange coupling ($K$) and spin-orbital coupling ($D$) as a function of the lattice parameter $a$ for different values of the Hubbard repulsion U.}
	\label{Fig_2}
\end{figure}

The multiple degrees of freedom renders VCl$_3$ a material realizing an effective 
$SU(4) = SU(2)_{\text{spin}}\times SU(2)_{\text{orbital}}$ model, 
similar to flat bands of twisted graphene multilayers \cite{PhysRevLett.128.227601,PhysRevX.9.041010,PhysRevLett.126.056803,PhysRevX.8.031087,PhysRevLett.127.026401,PhysRevB.108.045102,PhysRevX.11.041063}.
In the orbital degree of freedom, $SU(2)_{\text{orbital}}$ symmetry is broken into
a $U(1)_{\text{orbital}}$ symmetry due to the crystal lattice, in analogy with the
valley degree of freedom in twisted graphene multilayers.
Using first principles calculations we can map the effective Hamiltonian in the spin and
orbital degrees of freedom \cite{kugel1982jahn}, that takes the form

\begin{equation}
    H=-\frac{J}{2}\sum_{\langle ij \rangle} \vec{S}_i \cdot \vec{S}_j -\frac{K}{2}\sum_{\langle ij \rangle} \tau^z_i\tau^z_j - \frac{D}{2}\sum_{\langle ij \rangle}\vec{S}_i \cdot \vec{S}_j \tau^z_i\tau^z_j
    \label{H}
\end{equation}
where $\langle\rangle$ denotes first vanadium neighbors, $J$ is the isotropic Heisenberg-like coupling, $K$ is the anisotropic Ising-like coupling between pseudo-spins and $D$ represent the coupling between the spin and pseudo-spin variables. Since VCl$_3$ hosts two electrons in an high spin configuration, $S=1$, while the pseudo-spin can be described by an Ising coupling with $\tau^z=\frac{1}{2}$. Thus, $K$ capture the orbital exchange interaction while $D$ the coupling with the spin degree of freedom.  The schematic of this model is shown in Fig. \ref{Fig_1}c. We take in Eq. \eqref{H} that there is a strong
easy axis for the pseudospin,
associated with the explicit breaking of $SU(2)_{\text{orbital}}$ due to the lattice. 
This assumption is justified by our first-principles calculation, which converges to the same orbital configuration shown in Fig. \ref{Fig_1}b, even when the $d$-density matrix is initialized with rotated axes. Depending on the values of $J$, $K$ and $D$ different spin-orbital orders emerge from this model  \cite{corboz2012spin,tokura2000orbital,van1999double}. 
By ab-initio calculation, we calculated the effective parameters present in the Hamiltonian \eqref{H} through the stabilization of four possible ground states: (I) FO-FM, (II) FO-AFM, (III) AFO-FM, (IV) AFO-AFM (see Fig. \ref{Fig_2}a), where FM and AFM stands for ferromagnetic and antiferromagnetic respectively. 
We treat equation \eqref{H} in the classical approximation, i.e. describing spins $\vec{S}$ as dimensionless classical vectors of length $S$ in the sphere and considering $\tau^z_i\tau^z_j=\pm 1$. 
We denote the corresponding ground state energies as $\epsilon_{FO,FM}$, $\epsilon_{FO,AFM}$, $\epsilon_{AFO,FM}$ and $\epsilon_{AFO,AFM}$. The spin-orbital model allows to write the energy per unit cell (honeycomb lattice 2 V atoms)  of the different configurations as \cite{note4}
$
    \epsilon_{FO,FM}= -3JS^2-3K\textsuperscript{2}-3D\textsuperscript{2}S^2+E_0
    \label{1}
$.
where $E_0$ is a constant energy term. In order to determine $J$, $K$ and $D$, we use the ground state energies for these 4 configurations as obtained from our DFT calculations. The obtained results are reported in Fig. \ref{Fig_2}b-c-d as a function of the lattice parameter $a$ and for different values of the Hubbard U. We note that the theoretical lattice constant at the DFT+$U$ level is $a=6.24$ \AA  to be compared with available experimental lattice parameter for the bulk is $a=6.01$ \AA 
 \cite{Klemm:1947}. In general, we find $J>K>>D$, however, depending on the strain, the spin-orbital coupling $D$ can be enhanced by one order of magnitude. 
With 
$a=5.80$ \AA, close to the experimental bulk lattice parameter, $D=-0.27$ meV independently on the U value. It is worth noting that the exchange orbital coupling $K$ is highly dependent on the value of U. This aligns with the scenario where correlation effects drive the orbital ordering. Specifically, higher U values result in stronger orbital coupling, while in the limit of small U, the orbital coupling is quenched. Another important feature is the dependence of the spin exchange coupling $J$ on lattice strain,
where at some strains the coupling switches from ferromagnetic to antiferromagnetic.
This phenomenology may account for the significant impact of the substrate on the magnetic order, as recently demonstrated in Ref. \cite{deng2024ferroelectricity}. This allows for strain engineering of the magnetism, which in this material is coupled with the orbital and consequently the charge degrees of freedom. Finally, we note how for $a=5.9$ \AA, $J$ is almost quenched while $D$ is enhanced. In this regime, controlling the orbital configuration permits to switch from ferromagnetic to antiferromagnetic phase and viceversa. Thus, the  first principle  mapping on the Hamiltonian in Eq. \eqref{H} reveals a sizable coupling between spin and orbital degrees of freedom in VCl$_3$. 

\begin{figure*}[ht]
        \centering
	
        \includegraphics[width=1.9\columnwidth]{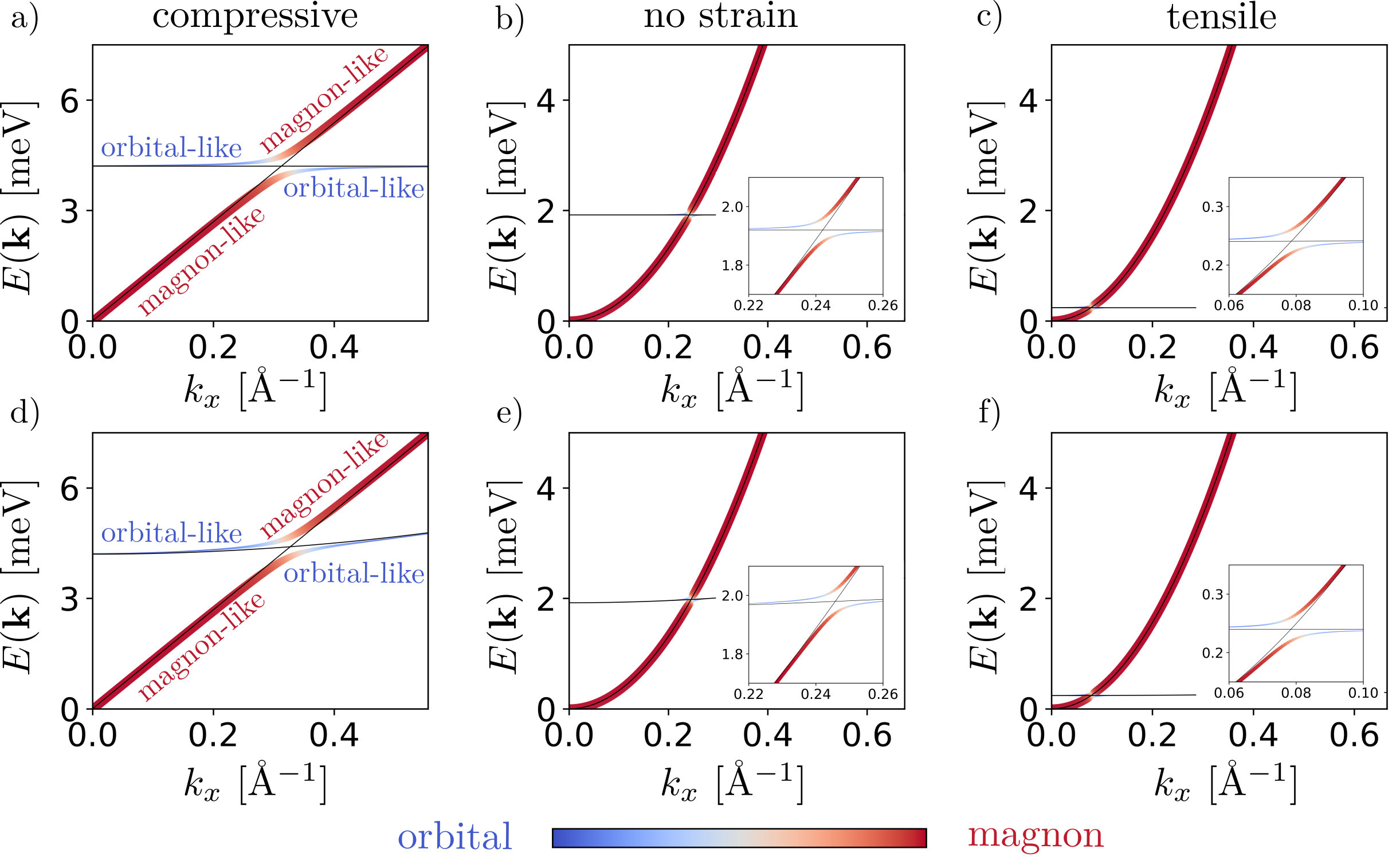}
	\caption{Magnon-orbital excitation as a function of the strain. In particular in a), b) and c) the excitation spectra are reported for the coupling values reported in Fig. \ref{Fig_2}. In  d), e) and f) a $\textbf{k}$ dispersion for the orbital excitation is assumed ($\omega_o(\textbf{k})$)), representing the coupling with other orbitals (see main text). The color bar represent the excitation character and the linewidth is proportional to the projection on magnon manifold. The black thin lines represent the magnon and orbital dispersion without hybridization ($D=0$).}
	\label{Fig_3}
\end{figure*} 
The magnetic and orbital orders breaking gives rise to magnetic and orbital excitations which can be studied using the the Hamiltonian in Eq. \eqref{H}. 
Considering only the spin degree of freedom, gapless magnons will arise due to the continuous symmetry of the spin sector. Due to the 2D nature of the magnetism, magnetic anisotropy is necessary for magnetic order to occur. We do not include it in our model, as it does not influence the magnon dispersion aside from opening a small magnon gap \cite{Lado2017}. Orbital excitations will have a gapped spectra due to the Ising-like interaction considered. To study elementary excitation we can consider Hamiltonian \eqref{H} as composed $H=H_s+H_o+H_{int}$, where $H_s$ is the only spin Hamiltonian, $H_o$ the orbital one and $H_{int}$ is the interaction term proportional to $D$. $H_s$ and $H_o$ are studied in terms of linear spin waves (LSW) and linear orbital waves (LOW) \cite{sw1_1989modified,sw2_2012interacting,wohlfeld2009orbitally}. We perform first an Holstein-Primakoff transformation introduce bosonic operators $b_i$ and $\alpha_i$ ($\tau^z_i=\tau^z-b^{\dagger}_ib_i$, $S^z_i=S-\alpha^{\dagger}_i\alpha_i$, $S^+_i=\sqrt{2S}\alpha_i$, $S^-_i=\sqrt{2S}\alpha^{\dagger}_i$ for the ferromagnetic case), then a Bogoliubov transformation gives the magnon dispersion and introduce magnon creation and annihilation operators $a^{\dagger}_i$,$a_i$. When the spin-orbital coupling is considered via $H_{int}$, the full Hamiltonian $H$ becomes biquadratic and a decoupling \cite{dec1_2006spin,dec2_2000quantum} is performed. By introducing a spin-orbital hybridization function and neglecting the anomalous terms, we can write down the effective Hamiltonian $H_{eff}$ as

\begin{equation}
\begin{split}
    H_{eff}= \sum_{\nu,\textbf{k}} \omega_{\nu}(\textbf{k}) a^{\dagger}_{\textbf{k},\nu}a_{\textbf{k},\nu} +\sum_{\textbf{k}} \omega_o(\textbf{k}) b^{\dagger}_{\textbf{k}}b_{\textbf{k}} \\ 
   +\sum_{\nu,\textbf{k}} \gamma^{\nu}(\textbf{k})a^{\dagger}_{\textbf{k},\nu}b_\textbf{k} + h.c.
\end{split}
    \label{H_eff}
\end{equation}

where $a^{\dagger}_{\textbf{k},\nu}$ $b^{\dagger}_{\textbf{k}}$ are bosonic operators which create a magnon with energy $\omega_{\nu}(\textbf{k})$ in the magnon band $\nu$ and an orbiton with energy $\omega_o(\textbf{k})$ respectively. The spin-orbital hybridization function 
 depends on the magneto-orbital coupling as $\gamma^{\nu}(\textbf{k})\sim D\langle a_{\textbf{k}}b^{\dagger}_{\textbf{k}}\rangle$. $\omega_o(\textbf{k})$ is the analogous of $\omega_{\nu}(\textbf{k})$ for the magnons, but in our Hamiltonian \eqref{H}, due to the Ising-like interaction between pseudo-spins, $\omega_o(\textbf{k})=6K$. If we consider the possibility of orbital in-plane rotation or a bond-dependent coupling, $\omega_o(\textbf{k})$ could depend on $\textbf{k}$ resulting in a orbital excitation dispersion. Finally, the magneto-orbital coupling also renormalizes the exchange spin and orbital interaction, as shown by rewriting Eq. \eqref{H} in the following form $H=\sum_{\langle i,j \rangle} J_{ij}(\vec{S}_i \cdot \vec{S}_j+1)$ where $J_{ij}=J_{ij}(\tau^z_i,\tau^z_j)$ \cite{dec3_1997spin}.

 The Hamiltonian \eqref{H_eff} is solved by expanding the Hilbert space of the spin states (first term in Eq. \eqref{H_eff}) to include the pseudo-spins (second term in \eqref{H_eff}) and their hybridization with spins (third term in \eqref{H_eff}). Since $J>K$ we expand the magnon and orbital dispersion near the $\Gamma$ point. In this limit, ferromagnetic dispersion is parabolic with an effective mass $m_{eff,S}=\frac{1}{4}JSa^2$ while antiferromagnetic dispersion is linear with an effective velocity $v_{eff}=\sqrt{\frac{3}{2}}JSa$ in an honeycomb lattice. 
The results of the entangled magnon and orbital
spectrum are summarized in Fig. \ref{Fig_3} as a function of the strain.
We first consider the case of the Hamiltonian \eqref{H}, in which orbitals cannot rotate \ref{Fig_3}a-b-c.
In absence of spin-orbital hybridization ($D=0$), the spectra consist into two different dispersions,
$\omega_{\nu}(\textbf{k})$ and $\omega_o(\textbf{k})$, which do not interact.
In the presence of magneto-orbital coupling ($D \neq 0$), a gap appears in the magnon spectrum, with its amplitude proportional to the coupling. Moreover, magnons and orbitons hybridize at the crossing points opening a gap giving rise to magnon-orbiton excitations (Fig. \ref{Fig_3}). Thus, the multicomponent order gives rise to both orbiton excitations as the one detected in Ref. \cite{saitoh2001observation} and hybridized magnon-orbiton excitations which, at the best of our knowledge, have never been experimentally observed. 
In Fig. \ref{Fig_3}d-e-f we consider the general case in which the interaction between orbital is of the type $\vec{\tau}^{(\gamma)}_i K_{ij}^{(\gamma)} \vec{\tau}^{(\gamma)}_j$, i.e. the pseudo-spin interaction allows both in-plane rotation and bond-dependent ($\gamma$) coupling. 
This leads to the emergence of an orbital dispersion $\omega_o(\textbf{k})$ \cite{dec1_2006spin,dec2_2000quantum,dec3_1997spin}. In this case the dispersion is not linear due to the strong anisotropy that describes the pseudo-spin variable. 
In order to verify if the energy gap is preserved including orbital dispersion, we consider an
orbiton dispersion with an effective mass $m_{eff,O}=K_{\perp}\tau^za^2$ where $K_{\perp} \simeq \frac{1}{16} K$ as estimated in Ref. \cite{dec3_1997spin}. The main consequence of $\omega_o(\textbf{k})$ is the tilting of the magnon bands at the points where the magnon and orbiton bands intersect, still showing the presence of a gap. This effect becomes more pronounced as the orbital coupling increases (see Fig. \ref{Fig_3}d). 
Another consequence of the general pseudo-spin interaction, which we do not explore in this work but is worth mentioning, is the emergence of a Kitaev-like interaction between spins in the limit of strong SOC \cite{KK_Kitaev_1_2018,KK_Kitaev_2_2023}. This is not the case for VCl$_3$ due to the lighter halide, but it could be the case for VI$_3$ if entangled magneto-orbital ordered phases can also be stabilized in this compound. 
Finally, we show how the amplitude of the gap and the strength of the hybridization can be tuned by in-plane strain. In particular, for a relatively large $6\%$ compressive strain (Fig. \ref{Fig_3}a-d) the system is an antiferromagnet with a very large magneto-orbital coupling ($D=-0.27$ meV) due to the increased hybridization, but only a $0.2\%$ tensile strain strongly decreases the orbital exchange coupling eventually causing it to disappear \footnote{For $a>6.4 \text{\AA}$ our first principles calculation cannot converge both FO and AFO phases.}.

Here we show that VCl$_3$ develops multiferroic order in both the orbital and spin degrees of freedom, giving rise to intertwined magnon and orbital excitations. Through first-principles calculations, we establish the appearance of strain-dependent ferro- and antiferro-orbital ordered phases in VCl$_3$, coexisting with the magnetic state. Based on first-principles methods, we derived a low-energy Hamiltonian that accounts for the magnon and orbital excitations, including the effect of magneto-orbital coupling between both degrees of freedom.
The magneto-orbital coupling gives rise to magneto-orbiton excitations stemming from the multicomponent order, establishing an exotic excitation that can be probed in this monolayer material. We show that magneto-orbital coupling, orbital order, and magnetic order are tunable with strain, making VCl$_3$ a potential platform for magneto-orbital straintronics. In particular, inhomogeneous strain, such as that present in twisted heterostructures, is expected to give rise to moir\'e domains with different magneto-orbital ordering and magneto-orbital dynamics, providing a new playground in moir\'e matter.
Our results establish the deep interplay between orbital and magnetic ordering in VCl$_3$, presenting a paradigmatic example of a multicomponent ordered phase in van der Waals materials.

\textit{Acknowledgements-} G.P. acknowledges the European Union-NextGenerationEU under the Italian Ministry of University and Research (MUR) National Innovation Ecosystem Grant No. ECS00000041 VITALITY-CUP E13C22001060006 for funding the project. G.P. and L.C. acknowledge support from CINECA Supercomputing Center through the ISCRA project. J.L.L. and A.O.F. 
acknowledge the computational resources provided by the Aalto Science-IT project
and the financial support from the Academy of Finland Projects Nos. 331342, 358088, and 349696,
the Jane and Aatos Erkko Foundation,
and the Finnish Quantum Flagship.

\bibliography{bibliography_new.bib}

\begin{thebibliography}{72}
\bibitem [{not({\natexlab{a}})}]{note}%
  \BibitemOpen
  \href@noop {} {} ({\natexlab{a}}),\ \bibinfo {note} {the $d_{z^2}$ in global coordinates is the $a_{1g}$ orbital in a trigonal symmetry, while the $d_{xy} \sim e'_{g,1}$ and $d_{x^2-y^2} \sim e'_{g,2}$ orbitals are a good representation of the $e'_g$ manifold.}\BibitemShut {Stop}%
\bibitem [{not({\natexlab{b}})}]{note4}%
  \BibitemOpen
  \href@noop {} {} ({\natexlab{b}}),\ \bibinfo {note} {we only write down FO-FM phase for the sake of brevity}\BibitemShut {NoStop}%
\end{thebibliography}

\end{document}